\documentclass[fleqn,10pt]{wlscirep}
\usepackage[utf8]{inputenc}
\usepackage[T1]{fontenc}

\usepackage{times}
\usepackage{hyperref}       % hyperlinks
\usepackage{url}            % simple URL typesetting
\usepackage{booktabs}       % professional-quality tables
\usepackage{amsfonts}       % blackboard math symbols
\usepackage{nicefrac}       % compact symbols for 1/2, etc.
\usepackage{microtype}      % microtypography
\usepackage{lipsum}
\usepackage{float}
\usepackage{multicol}
\usepackage{multirow}
\usepackage{tabularx}
\usepackage{graphicx}
\usepackage{xcolor}
\usepackage{colortbl}
\usepackage{authblk}

\usepackage{amsmath}
\usepackage{amsthm}
\usepackage{enumerate}
\usepackage{enumitem}

\setlist{leftmargin=5.5mm}
\hypersetup{
    colorlinks=true,
    linkcolor=black,
    filecolor=blue,      
    urlcolor=blue,
}
\usepackage[textwidth=0.1in,textsize=footnotesize]{todonotes}
\usepackage{xspace}

\usepackage{lineno}

\title{FoldExplorer: Fast and Accurate Protein Structure Search with Sequence-Enhanced Graph Embedding}

\author[1]{Yuan Liu}
\author[1,*]{Hong-Bin Shen}
\affil[1]{Institute of Image Processing and Pattern Recognition, Shanghai Jiao Tong University, and Key Laboratory of System Control and Information Processing, Ministry of Education of China, Shanghai, China}
\affil[*]{hbshen@sjtu.edu.cn}

\begin{abstract}
The advent of highly accurate protein structure prediction methods has fueled an exponential expansion of the protein structure database. Consequently, there is a rising demand for rapid and precise structural homolog search. Traditional alignment-based methods are dedicated to precise comparisons between pairs, exhibiting high accuracy. However, their sluggish processing speed is no longer adequate for managing the current massive volume of data. In response to this challenge, we propose a novel deep-learning approach FoldExplorer. It harnesses the powerful capabilities of graph attention neural networks and protein large language models for protein structures and sequences data processing to generate embeddings for protein structures. The structural embeddings can be used for fast and accurate protein search. And the embeddings also provide insights into the protein space. FoldExplorer demonstrates a substantial performance improvement of 5\% to 8\% over the current state-of-the-art algorithm on the benchmark datasets. Meanwhile, FoldExplorer does not compromise on search speed and excels particularly in searching on a large-scale dataset.
\end{abstract}
\begin{document}

\flushbottom
\maketitle

\thispagestyle{empty}

\section*{Introduction}
The exploration of protein homologs is a crucial method for protein analysis. Limited by the number of known protein structures, the identification of protein homologs has primarily relied on sequence alignment methods over the years. Various methods based on sequence similarity search like BLAST~\cite{altschul1990basic}, HH-suite~\cite{steinegger2019hh}, and MMseqs~\cite{steinegger2017mmseqs2}, have been established and widely used. Sequence comparison does help in many situations and achieves significant results. It seems not always effective because of the “twilight zone~\cite{doolittle1986urfs}” in protein space – in which proteins with similar structures or functions but low sequence identity. Detecting remote homologs through sequence search poses a significant challenge.

An inspiring development in protein structure prediction brings a huge shock to structural biology. Some structural prediction methods, such as AlphaFold2~\cite{jumper2021highly}, have achieved or approached experimental levels in terms of prediction accuracy. They also established a database for the predicted structure~\cite{varadi2022alphafold}. Hundreds of millions of predicted structures have greatly enriched the protein structure database and provided data support for research on structure comparison methods. Compared to sequence alignment, structure comparison can discover more significant structural homologs~\cite{vanni2022unifying}. Some algorithms based on structure comparison have been developed. TM-align~\cite{zhang2005tm} executes a residue-to-residue alignment based on structural similarity using heuristic dynamic programming iterations, while Dali~\cite{holm2020using} uses distance matrix alignment for structural comparison. These methods all have high accuracy, but their speed is very slow. For example, when using TM-align to search in a database with a million structures for a single query structure, it will take several days on one CPU core. If a large number of searches are to be performed, the time and computing resources consumed are huge. 

Therefore, some studies have used methods that are alignment-free to speed up the comparison process. They represent proteins as a vector and calculate the similarity of protein structures through the distance between vectors. These methods can be roughly divided into two categories based on the way they obtain structure representation vectors. One is based on the traditional mathematical methods. These methods construct complex mathematical models or utilize statistics knowledge to extract features from protein structures. For example, Scaled Gaussian Metric (SGM)~\cite{rogen2003automatic} extracts Gaussian invariants from the protein backbone based on knot theory to represent the protein structure as a 30-dimensional vector and uses the Euclidean distance between vectors as a measure of the distance between protein structures. Secondary Structure Element Footprint (SSEF)~\cite{zotenko2006secondary} counts the frequency of occurrence of secondary structure triples to represent the protein structure as a 1500-dimensional vector and uses the Pearson correlation coefficient as the measure of distance. Similar methods include Fragbag~\cite{budowski2010fragbag}, etc. The other is based on deep learning methods, such as DeepFold~\cite{liu2018learning}, GraSR~\cite{xia2022fast}, etc. They train deep neural networks to learn some inherent mechanisms of protein structure. DeepFold develops a convolutional neural networks (CNNs) model to extract structural motif features from protein distance maps. GraSR uses graph neural networks (GNNs) to characterize protein structure and trains using a contrastive learning framework, achieving better performance.

Foldseek~\cite{van2023fast} shone brilliantly due to its distinguished performance in protein structure search tasks. Foldseek has a very fast speed while achieving high search accuracy and sensitivity. It combines the advantages of alignment methods and alignment-free methods. Foldseek constructs a 3Di alphabet by training a VQ-VAE architecture network. Protein structures are discretized into 3Di sequences. This transformation converts structural alignment into the alignment of 3Di sequences, significantly improving the comparison speed.

However, these existing methods face a common challenge—difficult to strike a balance between accuracy and time efficiency when dealing with large-scale datasets. For example, representation-based approaches, while achieving faster processing speeds, struggle to ensure high accuracy and may discard a substantial number of potential homologs. In contrast, Foldseek maintains a higher accuracy level and also significantly boosts the speed of alignment-based methods, but this is only in comparison to slower methods like TM-align, Dali, and CE~\cite{shindyalov1998protein}. According to our testing, when conducting comprehensive comparisons and searches across all 542,378 SwissProt structures predicted by AlphaFold2, Foldseek requires over 400 hours with 16 CPU threads.

In this paper, we propose a novel protein structure search framework FoldExplorer. FoldExplorer employs a sequence-enhanced graph embedding approach to represent protein structures. It harnesses the strengths of pre-trained large-scale protein language models for protein sequences while capitalizing on GNNs' capabilities to learn geometric data. GNNs are used to extract information from protein structures. And we also integrate the powerful large-scale language model ESM2~\cite{lin2022language} into our pipeline, leveraging its robust sequence analysis capabilities to augment structural representation, ultimately leading to improved performance. The results show that FoldExplorer outperforms SOTA methods in both ranking and classification tasks. Additionally, it exhibits a faster processing speed.

\begin{figure}[hp]
    \centering
    \includegraphics[width=0.9\textwidth]{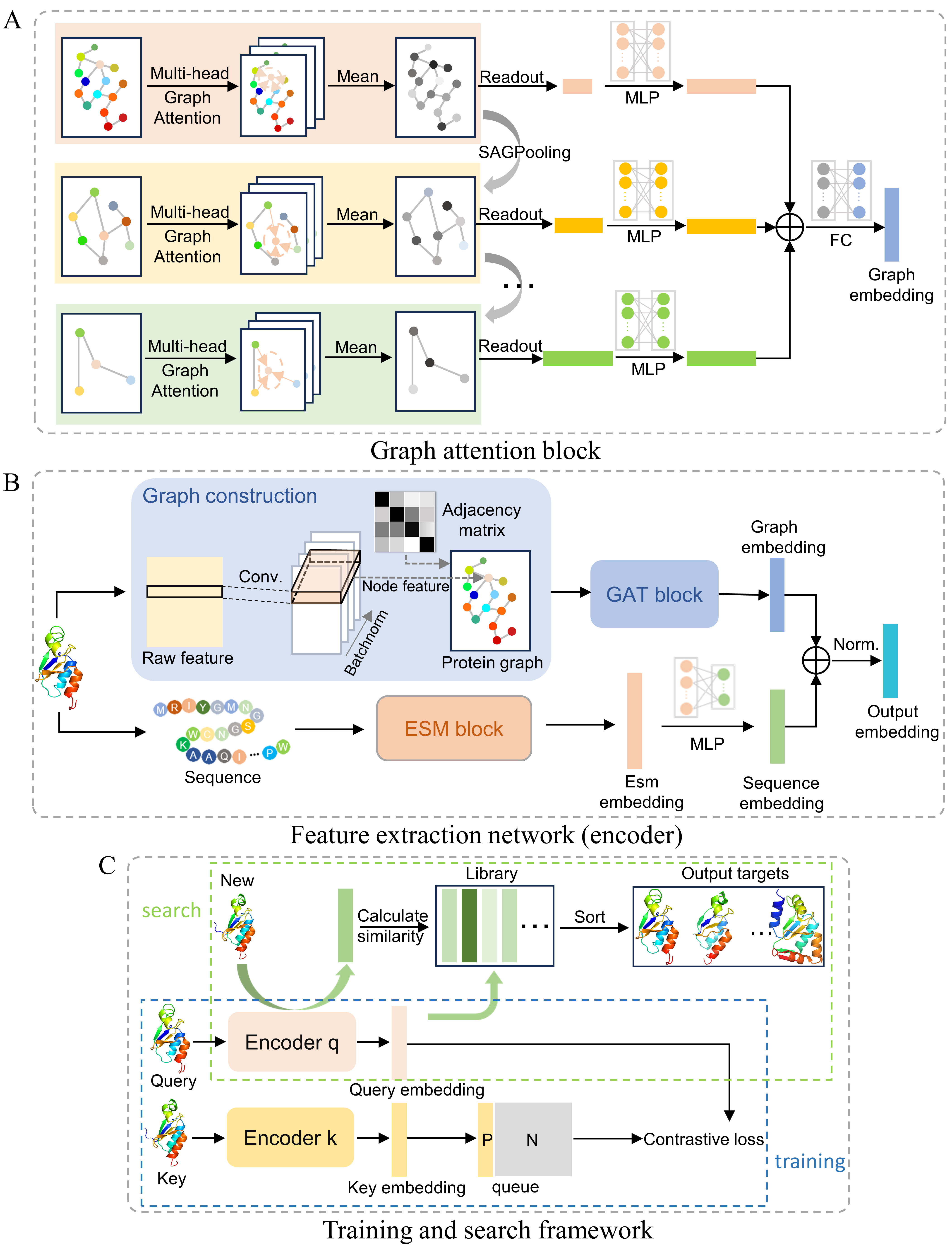}
    \caption{\label{fig:1}\textbf{Pipeline of Foldexplorer.}
    \textbf{A}.  Graph-based encoder.
    \textbf{B}.  Feature extraction network.
    \textbf{C}.  Training and search framework.
    }
\end{figure}

\section*{Materials and Methods}
\paragraph{Benchmark dataset.}
This study uses SCOPe 2.07~\cite{fox2014scope, chandonia2022scope} (published in March 2018) as the benchmark dataset. To remove redundancy and eliminate the effect of sequence similarity on the structural comparison, we only use the subset with less than 40\% sequence identity, a total of 14323 protein structure domains. To facilitate a fair comparison with the baseline methods, we used the same data screening method as they did, and finally obtained 13265 protein domains. We divided these data into five equal parts for 5-fold cross-validation, and all structures are used to build a library for protein similarity structure search. 

In addition, we construct an independent test set from the Protein Data Bank (PDB)~\cite{sussman1998protein} database. We randomly collected proteins with release dates between October 1, 2021 and May 1, 2023 and with high-resolution PDB format structure data stored in the PDB database. We selected the chain with domain annotations and intercepted the corresponding structural domain from the PDB file. Afterward, use the cd-hit~\cite{li2006cd} tool to remove sequential redundancy, both within the test set and between the test set and the SCOPe dataset, with a cutoff of 40\%. Finally, 1076 protein structures were obtained as a non-redundant independent test set, named indDomain.

There is no overlap among the two benchmark datasets mentioned above.

\paragraph{Sequence-enhanced structure representation.}
We utilize a sequence-enhanced structural representation network to generate a 512-dimensional vector as a descriptor for each protein. It consists of two parts, one is a GAT~\cite{velickovic2017graph}-based encoder, and the other is a Sequential feature extractor based on large language models. Below we will introduce them separately.

\paragraph{Graph-based encoder.}
We use GAT to extract features from protein tertiary structures. GAT is one of the most powerful variants of GNNs. It has been widely used in many tasks related to molecular structure, such as protein function prediction~\cite{li2023deepgatgo}, drug-drug interaction prediction~\cite{nyamabo2021ssi}, and protein-ligand binding affinity prediction~\cite{li2021structure}. GAT can learn the weight of each neighboring node when aggregating messages with a self-attention mechanism, without the need for manually designed edge attribution, which reduces artificial participation and is more conducive to handling inductive tasks\cite{zhang2023gnngo3d}. 

Given a protein graph $G = (V, E)$, where $V$ is the node set and $E$ is the edge set. In our method, each $v_i \in V$ represents a residue, and if the Euclidean distance between the two residues $i$ and $j$ is less than 10 \AA, there will be an edge $e_{ij}$. All edges are undirected, and each node has a self-ring. We use a $(0, 1)$ matrix to represent the adjacency matrix $A$. If there is an edge between two nodes, the corresponding position of the adjacency matrix is $1$, otherwise, it is $0$. The $i-th$ node's feature is denoted as $h_i$. The GAT layer calculation process can be described by the formula as follows:

\begin{equation} \label{eq:GAT-1}
\centering
  \alpha_{i,j} = \frac{\exp(\sigma(\mathbf{a}^T[\mathbf{W}h_i || \mathbf{W}h_j]))}{\sum_{k\in \mathcal{N}(i)\cup\{i\}}\exp(\sigma(\mathbf{a}^T[\mathbf{W}h_i || \mathbf{W}h_k]))}  
\end{equation}

\begin{equation} \label{eq:GAT-2}
    h_i = \sum_{j \in \mathcal{N}(i) \cup \{i\}} \alpha_{i,j} \mathbf{W} h_j  
\end{equation}

Where $\mathbf{W}$ is the shared weight matrix and $\alpha$ is the attention coefficients, $\sigma(\cdot)$ is the activation function, here we use LeakyReLU. When there are multiple attention heads, each has an independent weight matrix and output, and finally the mean of all outputs as the GAT convolution layer output. Each layer’s output is not only input to the next layer through a SAGPooling layer but also reaches the final output layer through global average pooling (GAP) and skip connection. We want to use this method to extract key nodes and condensed features step by step and also prevent gradient vanishing caused by stacking too many layers. The final output of the graph-based encoder we called graph embedding is calculated as follows:

\begin{equation} \label{eq:SAGpooling}
    G^{(l+1)} = SAGPooling(G^{(l)})
\end{equation}

\begin{equation} \label{eq:graph embedding}
    Graph\ embedding = \sum_{i=1}^n MLP(GAP(G^{(i)}))
\end{equation}

\paragraph{Feature extraction network.}
As shown in Fig.~\ref{fig:1}\textbf{B}, to better represent protein structure, we use two different levels of information: tertiary structure and primary structure, and design feature extraction modules respectively. The tertiary structure is encoded in a protein graph. Each residue is regarded as a node in the graph, and the initial node features are the same as GraSR, that is, derived from the C$\alpha$ atom coordinates, including distance-based and angle-based features. They are rotational and translational invariance~\cite{xia2022fast}. The adjacency matrix is obtained according to the distance between the residue C$\alpha$, if the distance is less than the threshold, there is an edge between the two nodes, and the threshold is taken as 10\AA.

We use pre-training models to extract features from protein sequences. ESM is one of the most powerful large-scale protein sequence language models. Firstly, we use the ESM-2 model with 650 million parameters pre-trained on Uniref50 to encode the protein sequences and obtain embeddings at the protein level. Then, we use an MLP for fine-tuning to get sequence embeddings. Finally, the sequence embeddings will be added to the graph embeddings calculated by the GAT block and then normalized as the final output embeddings.

\paragraph{Overall pipeline.}
Fig.~\ref{fig:1}\textbf{C} illustrates the training and search process. We use the contrastive learning framework to train the model. In contrastive learning, a point of view is that more negative samples in a batch is beneficial for training. Due to hardware resource constraints, we use MoCo's method to increase the number of negative samples while reducing memory overhead~\cite{he2020momentum}. Each batch contains a query set and a key set. Each query sample only forms a positive pair with the corresponding key sample, and negative pairs with other key samples. The query set and key set are encoded through two Siamese encoders $q$ and $k$ respectively. The two encoders are with the same structure but different parameters. The output of encoder $k$ in each batch goes into a queue to increase the number of negative samples. The contrastive loss function is InfoNCE Loss~\cite{he2020momentum}, the formula is as follows:

\begin{equation} \label{loss}
    L = -log \frac{exp({q \cdot k_+ / \tau})}
    {\sum_{i=0}^m exp({q \cdot k_i / \tau})}
\end{equation}

Where $q$ is the query embedding, $k_i$ is the $i-th$ embedding of the key embedding queue, $k_+$ is the positive key embedding, and $\tau$ is the temperature coefficient. All hyperparameters are set to the default values given by MoCo. The parameters of encoder $q$ are updated through gradient backpropagation, while the parameters of encoder $k$ are updated through momentum update as follows.

\begin{equation} \label{moco}
    \theta_k \leftarrow m \theta_k + (1-m) \theta_q
\end{equation}

In the search stage, we first encode the protein into an embedding vector through the encoder $q$. And then calculate the similarity among the structure library, the embeddings of the structure library have been calculated and stored in advance. Then, provide candidate sets based on calculated similarity. Finally, we sort the candidate set according to the calculated similarity score and output the targets.

\paragraph{Evaluation metrics.}
We adopt the methods from DeepFold and GraSR to determine positive and negative pairs. That is for protein A, calculate the TM-score~\cite{zhang2004scoring} with all other proteins in the dataset, where the maximum TM-score is denoted as $T_{MAX}$. If another protein B in the dataset has TM-score $T(A, B)$ not less than $0.9 * T_{MAX}$, then B is considered as a positive sample of A, also called a structural neighbor of A, otherwise, it is a negative sample.

For a fair comparison with the baseline methods, we also adopted the same 5-fold cross-validation training strategy and evaluation protocols defined by them. Like GraSR, we also apply a ranking task and a classification task. In the ranking task, we calculate the Area Under the Receiver Operating Characteristic (AUROC) and the Area Under the Precision-Recall Curve (AUPRC). Due to the significantly higher number of negative samples than positive samples, we are more concerned about AUPRC. The Top-k hit ratio evaluation metric~\cite{xia2022fast} defined in GraSR has also been adopted. It is defined as follows:

\begin{equation}
    Ratio_K = \frac{1}{N_q} \sum_{i=1}^{N_q}
    \frac{N_{hit}^i}{min(K, N_{nbr}^i)}
\end{equation}

Where $N_{hit}^i$ is the number of positive samples of $i-th$ query found by the algorithm in top $K$, $N_{nbr}^i$ is the truth number of positive samples. $N_q$ is the total number of queries. We calculate the ratio when $K$ = 1, 5, and 10. 

We tend to find positive samples that are most similar to queries. So we also calculated the fraction of the average TM-score of the Top $K$ samples to $T_{MAX}$, the formula as follows:
\begin{equation}
    Fraction_K = \frac{1}{N_q} \sum_{j=1}^{N_q}
    \frac{\sum_{i=1}^K T(q_j, top_j(i))}{K \cdot T_{MAX}(j)}
\end{equation}

Where $T_{MAX} (j)$ is the max TM-score of the $j-th$ query, $T(q_j,top_j (i))$ denotes the TM-score of the $j-th$ query and the corresponding $i-th$ target found by the algorithm.

In the classification task, the labels are obtained from the SCOPe dataset, and we evaluated them at four levels: class, fold, superfamily, and family. We calculate the F1-score and accuracy as the evaluation metrics at the class level. We also illustrate the precision-recall curve (PRC) on the other three levels to visualize the performance of each classifier. 

\section*{Results and Discussion}
In this section, we undertake a comprehensive comparative analysis of FoldExplorer against a selection of state-of-the-art structure search methodologies. This comparison encompasses a range of alignment-free and alignment-based approaches. 

\paragraph{Performance on the ranking task.}
Table~\ref{table1} presents the performance rankings of FoldExplorer alongside several baseline methods for 5-fold cross-validation using the SCOPe dataset. we have excluded Foldseek from our comparative analysis because of its utilization of the SCOPe dataset during its training process.
 
Since the highly imbalanced distribution of positive and negative samples, with a notably lower number of positive samples, our primary focus lies on the AUPRC, rather than the AUROC. In terms of AUPRC, FoldExplorer demonstrates a remarkable performance, surpassing the state-of-the-art method GraSR by a margin of 7.67\%. Additionally, when assessing the Top k Hit Ratio as defined by GraSR, FoldExplorer consistently outperforms its counterparts by 7.53\%, 6.58\%, and 5.16\%, respectively. This pronounced superiority underscores the effectiveness of FoldExplorer.
 
In the SCOPe dataset, the training and validation sets are from the same distribution. However, in typical protein structure search tasks, the query protein structures often stem from experimental data or computational predictions of novel structures. Therefore, we conducted analogous experiments on an independent test set, and the results are presented in Table~\ref{table2}. Because this set of structures is out of SCOPe distribution, the performance of other methods experiences a noticeable decline compared to their performance on SCOPe. Conversely, FoldExplorer's performance demonstrates no degradation. This indicates that FoldExplorer has a stronger generalization ability.

\begin{figure}[hp]
    \centering
    \includegraphics[width=1\columnwidth]{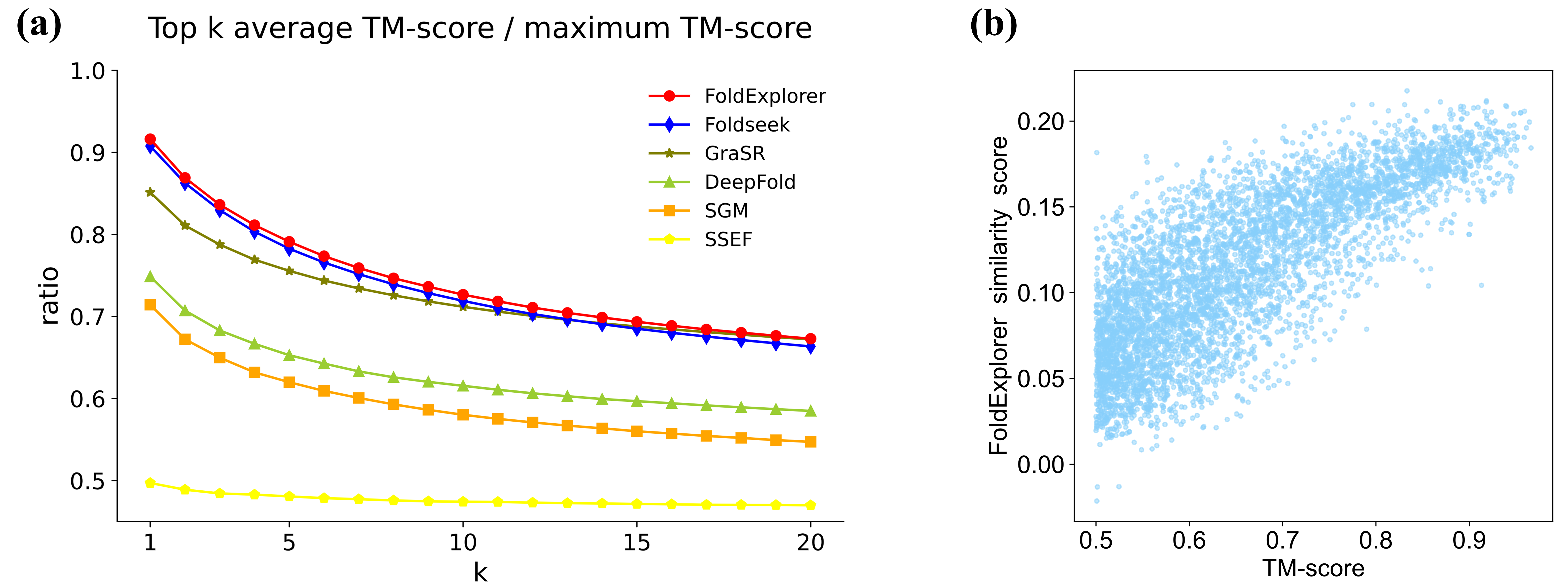}
    \caption{\label{fig:2}\textbf{Performance comparison on ranking task.} \textbf{(a)} Ratio of Top k average TM-score divided by the maximum TM-score. \textbf{(b)} Correlation of the similarity calculated by FoldExploror and TM-align}
\end{figure}

On the Top K hit ratio metric, FoldExplorer outperforms the alignment-free method GraSR by a significant margin of 36.6\%, 33.4\%, and 26.2\%, respectively. In comparison to the alignment-based method Foldseek, FoldExplorer also shows improvement of 5.19\%, 5.60\%, and 4.88\%. Furthermore, regarding the Top k TM-score metric, FoldExplorer's overall performance surpasses that of other baseline methods, as illustrated in Fig.~\ref{fig:2}\textbf{(a)}. These results indicate that, compared to other methods, FoldExplorer exhibits a stronger tendency to prioritize targets with higher TM-scores, which is what we expect in protein structure search tasks.

To evaluate the correlation between the similarity scores computed by FoldExplorer and the TM-score, we randomly sample 5000 structure pairs from the SCOPe dataset with TM-score greater than 0.5~\cite{xu2010significant}. The result is shown in Fig.~\ref{fig:2}\textbf{(b)}. The Spearman correlation coefficient between them reaches 0.7909. This indicates that, even though we do not treat TM-score as a label to fit, we only use it as a criterion for determining positive or negative samples, the similarity provided by FoldExplorer remains highly consistent with TM-align.

\begin{table}[ht]
\centering
\begin{tabular}{lccccc}
\toprule
\multirow{2}{*}{Methods} & \multirow{2}{*}{AUROC} & \multirow{2}{*}{AUPRC} & \multicolumn{3}{c}{Top K hit ratio} \\ \cline{4-6} 
                         &                        &                        & 1          & 5          & 10        \\ 
\midrule
SGM                      & 0.9224                 & 0.4537                 & 0.5562     & 0.5312     & 0.5553    \\
SSEF                     & 0.8423                 & 0.0381                 & 0.0838     & 0.058      & 0.061     \\
DeepFold                 & 0.9574                 & 0.4971                 & 0.6035     & 0.5659     & 0.5927    \\
GraSR                    & \textbf{0.9823}                 & \underline{0.6595}                 & \underline{0.7282}     & \underline{0.7101}     & \underline{0.74}      \\
FoldExplorer             & \underline{0.9758}                 & \textbf{0.7101}                 & \textbf{0.7830}      & \textbf{0.7568}     & \textbf{0.7782}    \\ 
\bottomrule
\end{tabular}
\caption{\textbf{Ranking performance on SCOPe dataset.} The best results are marked in bold and the second-best results with underline.}
\label{table1}
\end{table}

\begin{table}[ht]
\centering
\begin{tabular}{lcccccc}
\toprule
\multirow{2}{*}{Methods} & \multicolumn{3}{c}{Top K hit ratio} & \multicolumn{3}{c}{Top K TM-score fraction} \\ \cline{2-7} 
                         & 1          & 5          & 10         & 1             & 5            & 10           \\
\midrule
SGM                      & 0.3579     & 0.3470     & 0.3699     & 0.7143        & 0.6199       & 0.5802       \\
SSEF                     & 0.0498     & 0.0398     & 0.0437     & 0.4971        & 0.4807       & 0.4742       \\
DeepFold                 & 0.4096     & 0.3901     & 0.4209     & 0.7487        & 0.6528       & 0.6155       \\
GraSR                    & 0.5803     & 0.5642     & 0.5976     & 0.8513        & 0.7556       & 0.7118       \\
Foldseek                 & 0.7537     & 0.7128     & 0.7193     & 0.9077        & 0.7826       & 0.7190       \\
FoldExplorer             & \textbf{0.7928}     & \textbf{0.7527}     & \textbf{0.7544}     & \textbf{0.9162}        & \textbf{0.7911}       & \textbf{0.7266}  \\
\bottomrule
\end{tabular}
\caption{\textbf{Ranking performance on indDomain dataset.} The best results are marked in bold.}
\label{table2}
\end{table}

\begin{figure}[hp]
    \centering
    \includegraphics[width=1 \columnwidth]{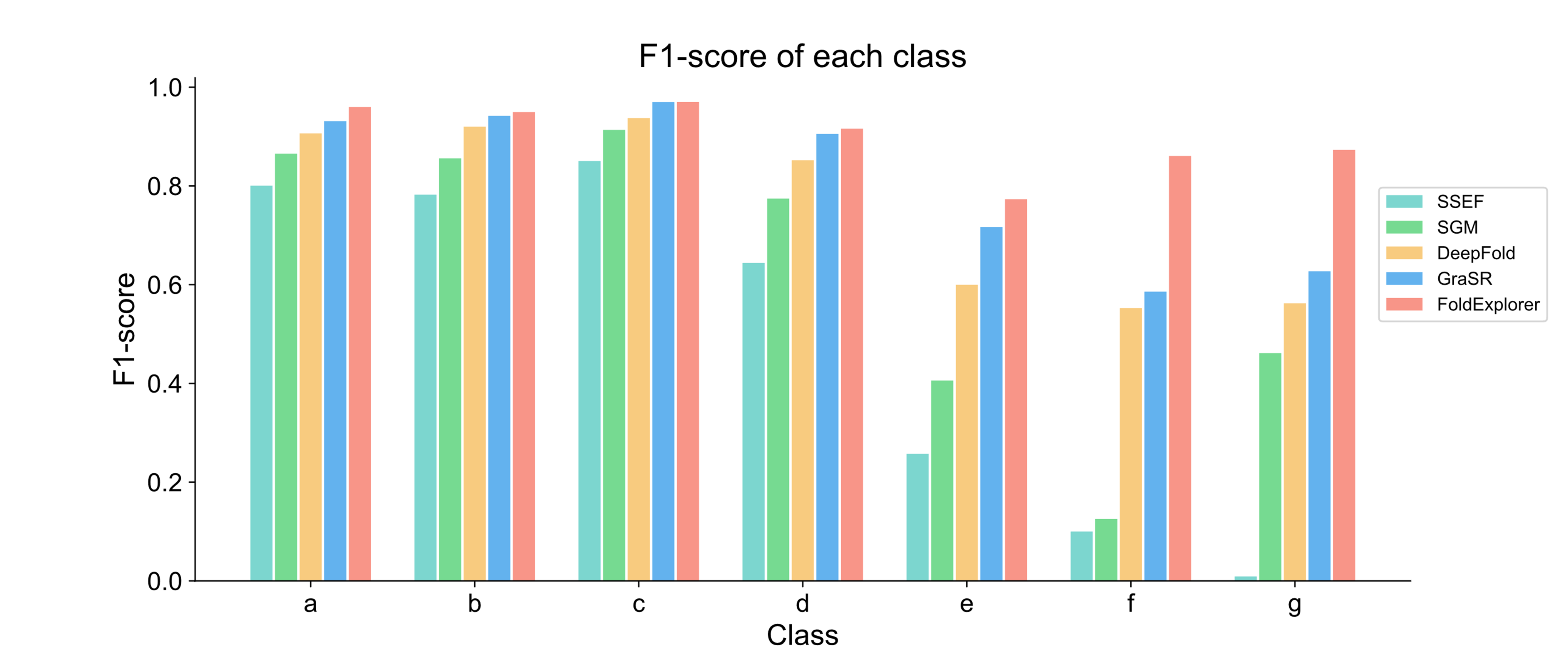}
    \caption{\textbf{F1-score of each class.}  a: All alpha proteins. b: All beta proteins. c: Alpha and beta proteins(a/b). d: Alpha and beta proteins(a+b). e: Multi-domain proteins(alpha and beta). f: Membrane and cell surface proteins and peptides. g: Small proteins.}
    \label{fig:3}
\end{figure}

\paragraph{Performance on the classification task.}
TM-score evaluates search results solely based on structural similarity. To comprehensively evaluate the performance of FoldExplorer, we also conducte classification experiments on the SCOPe dataset. SCOPe uses a tree-like structure to classify proteins, including four levels: class, fold, superfamily, and family. The dataset we use contains a total of 7 classes, 1150 folds, 1833 superfamilies, and 4245 families. At the class level, we compare FoldExplorer with other alignment-free methods. We feed the representation vectors obtained from each method into a logistic regression (LR) classifier and used 10-fold cross-validation to train and validate the classifier, respectively. Using the simplest classifier allows the difference in performance to come from the representation vectors rather than the classifier. We calculated the F1-score and accuracy for each class. The results are shown in Fig.~\ref{fig:3} and Table~\ref{tab:3}.

\begin{table}[hp]
    \centering
    \begin{tabular}{lcc}
\toprule
        Methods & Avg. F1-score & Accuracy  \\
\midrule
         SGM & 0.6289 & 0.8354 \\
         SSEF & 0.4920 & 0.7470 \\
         DeepFold & 0.7615 & 0.8887 \\
         GraSR & 0.8124 & 0.9258 \\
         FoldExplorer & \textbf{0.9004} & \textbf{0.9434} \\
\bottomrule
    \end{tabular}
    \caption{\textbf{Multi-class classification performance on SCOPe class level.} The best results are marked in bold.}
    \label{tab:3}
\end{table}

As depicted in Fig.~\ref{fig:3}, on the a, b, c, and d four classes, several other methods also achieve commendable F1-scores, with FoldExplorer showing a slight performance edge. However, for the e, f, and g three classes, FoldExplorer exhibits a distinct advantage, significantly outperforming its counterparts. FoldExplorer consistently excels across all seven classes, boasting an average F1-score that surpasses GraSR by 10.8\%. This underscores FoldExplorer's more comprehensive and discriminative representation of protein structures, effectively mitigating the limitations of other methods. Leveraging the feature vectors provided by FoldExplorer enhances the precision in classifying protein structures accurately.

\begin{figure}[ht]
    \centering
    \includegraphics[width=1 \columnwidth]{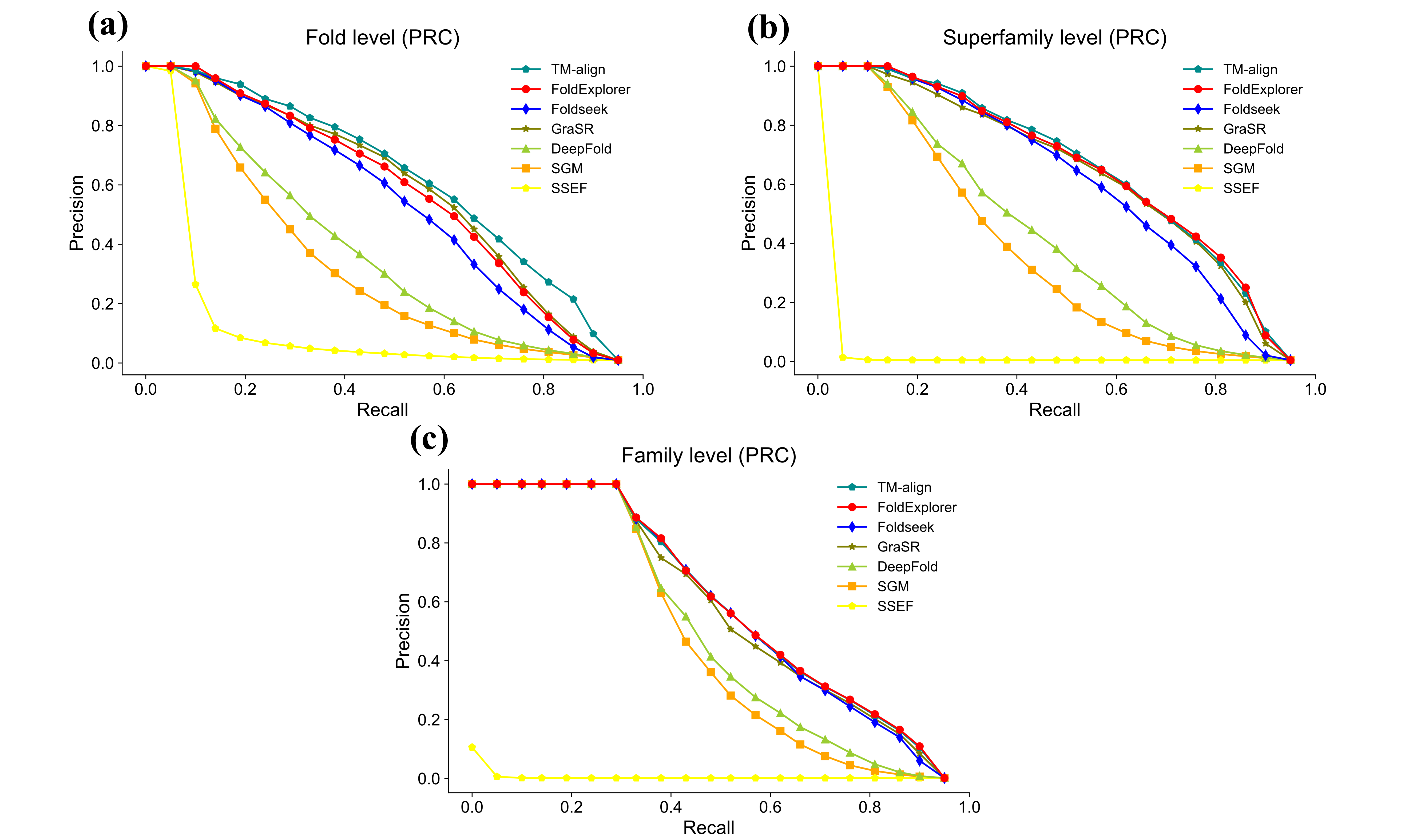}
    \caption{\textbf{Precision-Recall Curve (PRC) of the SCOPe dataset on different levels.}. \textbf{(a)} Fold level. \textbf{(b)} Superfamily level. \textbf{(c)} Family level.}
    \label{fig:4}
\end{figure}

Additionally, experiments are also conducted at the fold, superfamily, and family levels. We employ each method for an exhaustive all-versus-all search on the SCOPe dataset. During the search, targets are identified as true positives (TP) if they belong to the same category as the query, and false positives (FP) otherwise. Subsequently, we get precision-recall curves (PRC) as illustrated in Fig.\ref{fig:4}. FoldExplorer excels across these three levels, further emphasizing its propensity to rank targets belonging to the same category at the forefront during the search. In particular, when it comes to finer superfamily and family classifications, FoldExplorer has even achieved more accurate categorizations than TM-align. This inclination not only enhances search efficiency but also highlights FoldExplorer's outstanding performance in recognizing hierarchical features of protein structures, making it a powerful tool in the fields of protein classification and structure search.

\begin{figure}[htbp]
    \centering
    \includegraphics[width=1 \columnwidth]{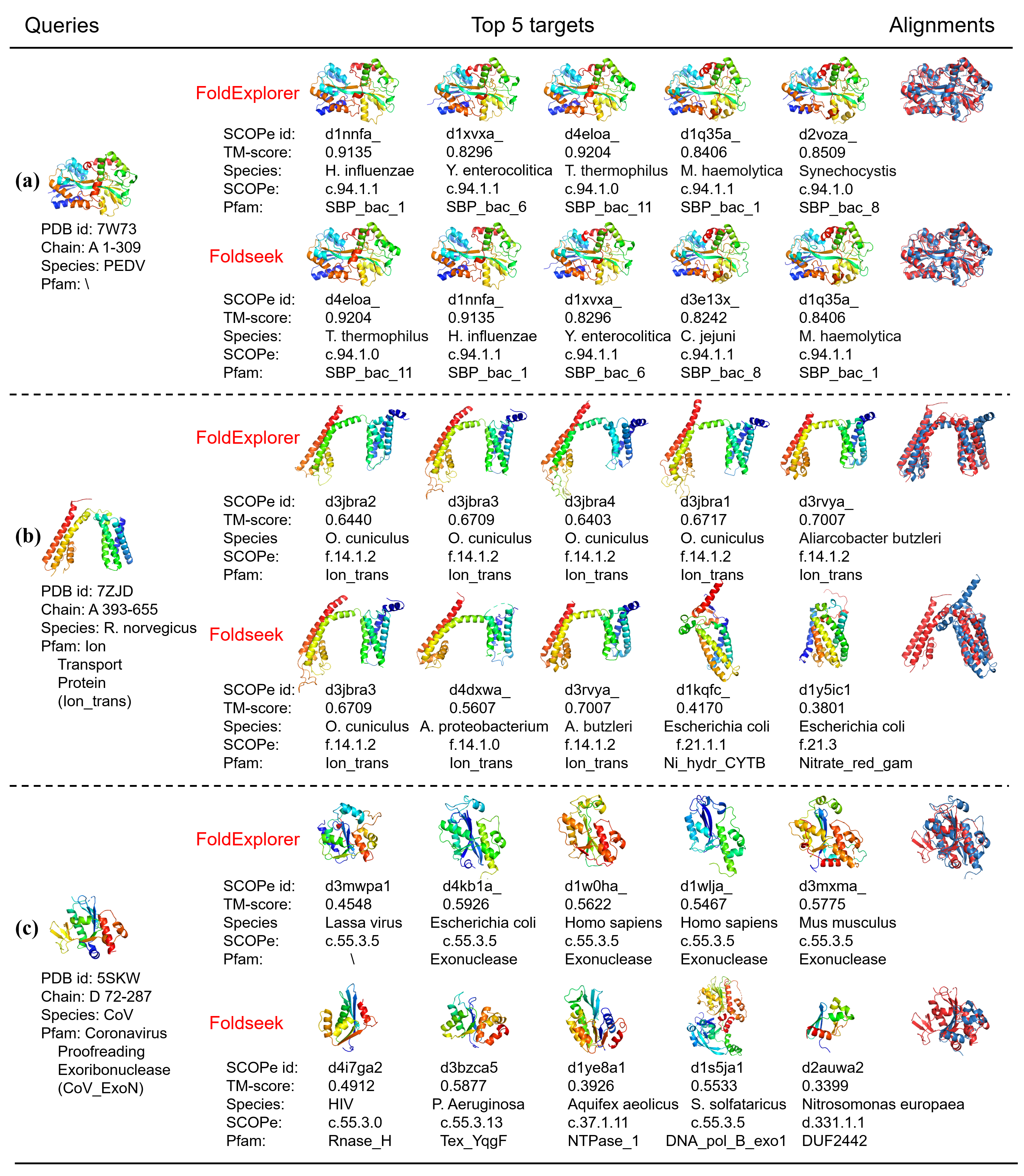}
    \caption{\textbf{Some search cases of FoldExplorer and Foldseek.}}
    \label{fig:5}
\end{figure}

\paragraph{FoldExplorer is more sensitive to remote homologs.}
Based on the statistic in Table ~\ref{table2} for the indDomain dataset, FoldExplorer's advantage over Foldseek may not seem pronounced. However, TM-score simply evaluates structural similarity and cannot capture deeper information about proteins, such as functions, etc. When conducting structural searches, our primary aim is to identify potential homologs. They may not have significant overlap in their overall structures. Instead, they may share common features within specific functional regions, or their shapes may exhibit overall similarities. The numerical calculation of TM-align struggles to capture this. From this perspective, FoldExplorer is more sensitive to homology compounds than Foldseek, surpassing even TM-align. We will elaborate on this with several examples in Fig.~\ref{fig:5}.

For each query, we present the top 5 targets identified by FoldExplorer and Foldseek, respectively. In some easy cases, the dataset contains proteins highly similar to the query, both FoldExplorer and Foldseek can locate corresponding targets, although their ranking may vary slightly, as shown in Fig.~\ref{fig:5}\textbf{(a)}. Achieving this is relatively straightforward and easy. However, when the similarity between query and targets in the dataset is not very high, as illustrated in Fig.~\ref{fig:5}\textbf{(b)}, the differences between FoldExplorer and Foldseek become more apparent. FoldExplorer focuses more on the overall protein structure, prioritizing proteins with similar overall shapes. Despite we input no information other than the structure, surprisingly, all the top 5 targets found by FoldExplorer are with the same Pfam~\cite{mistry2021pfam} annotation: ion transport protein as the query. But Foldseek seems to emphasize local structure in the last two targets. They only partially overlap with the query. In an extreme case, depicted in Fig.~\ref{fig:5}\textbf{(c)}, when the potential targets in the dataset have low similarity with the query (maximum TM-score below 0.6), FoldExplorer still produces satisfactory results. The query annotated by Pfam as “Coronavirus proofreading exoribonuclease (CoV\_ExoN)”, Among the 5 targets identified by FoldExplorer, 4 are annotated as Exonucleases. The remaining one lacks a corresponding Pfam annotation. In SCOPe, it belongs to the same family as the other 4. It is noteworthy that they originate from different species, ranging from humans to mice to viruses, and so on. This indicates FoldExplorer's ability to discover remote homologs. In contrast, Foldseek's top 5 targets do not exhibit this characteristic, although some targets have a relatively high TM-score. Foldseek tends to prioritize higher TM-scores only, whereas FoldExplorer can identify structures that are functionally closer, attributed to its utilization of sequence embedding.

These examples highlight that the targets identified by FoldExplorer possess relatively high TM-scores and, furthermore, demonstrate greater sensitivity to homologous relationships. This enhanced sensitivity aids in uncovering potential homologs through structural comparisons.

\begin{figure}[htbp]
    \centering
    \includegraphics[width=0.7 \columnwidth]{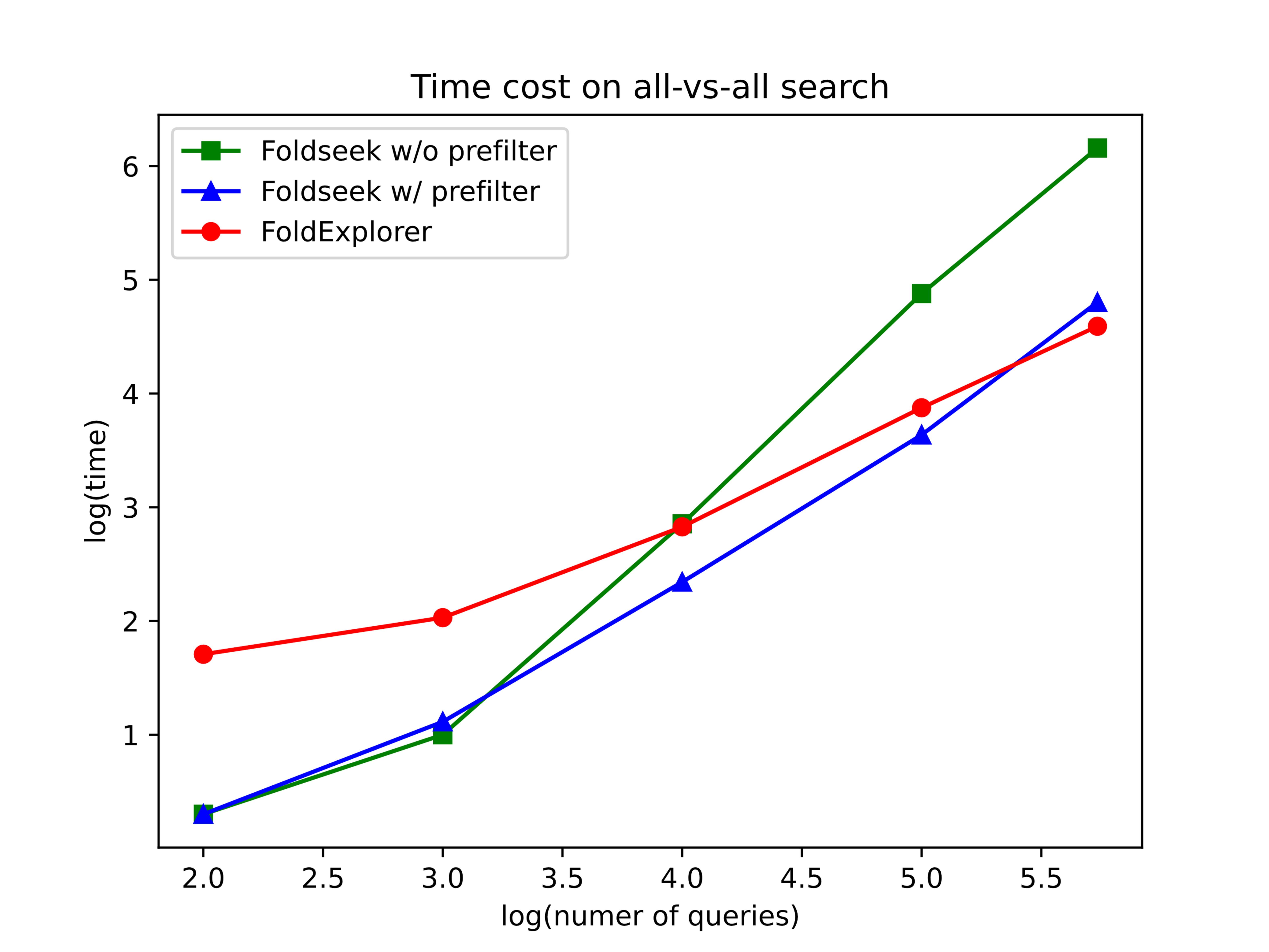}
    \caption{\textbf{Time cost comparison for FoldExplorer and Foldseek.}}
    \label{fig:6}
\end{figure}

\paragraph{FoldExplorer Excels in Computational Efficiency on Large Datasets
}
Currently, AlphaFold has predicted over 200 million protein structures. When conducting structure searches on such a massive dataset, the demands for both the speed and accuracy of the search algorithm are equally important. FoldExplorer performs exceptionally well on large-scale datasets. We conducted exhaustive all-vs-all search experiments using a platform with 2 * Intel Xeon Silver 4214R CPUs and an NVIDIA GeForce RTX 3090 GPU. Both methods employ 16 CPU threads. As shown in Fig.~\ref{fig:6}, the results indicate that when performing search tasks on the dataset of 542,378 structures from SwissProt, Foldseek without prefilter and with prefilter take 400 hours and 21 hours, respectively, whereas FoldExplorer only requires 10.8 hours. When the dataset is even larger, the performance gap becomes more pronounced. According to the experimental estimates, the time complexity of FoldExplorer is approximately $O(n)$, while Foldseek has a time complexity of approximately $O(n^2)$ without prefilter and approximately $O(n^{1.3})$ with prefilter. FoldExplorer demonstrates superior performance on large-scale datasets.

\paragraph{FoldExplorer learns the distribution of the protein space.}
In the context of protein structure search tasks, we can imagine all protein structures as residing in a high-dimensional space. 

Alignment-based methods compare a query with others, providing only the concept of distances between structures. The specific spatial positions of structures are uncertain, making it difficult to understand the distribution of various structure types in space. However, representation-based methods approach this problem from another perspective. They map protein structures into a latent space, where each structure's position is determined. Then, by defining a distance metric in this latent space, they characterize the proximity of proteins in structural space. A robust representation method should map the structures in a way that corresponds to the original structural space, meaning that protein pairs neighbors in the original space should also be closed in the latent space.

\begin{figure}[htbp!]
    \centering
    \includegraphics[width=1 \columnwidth]{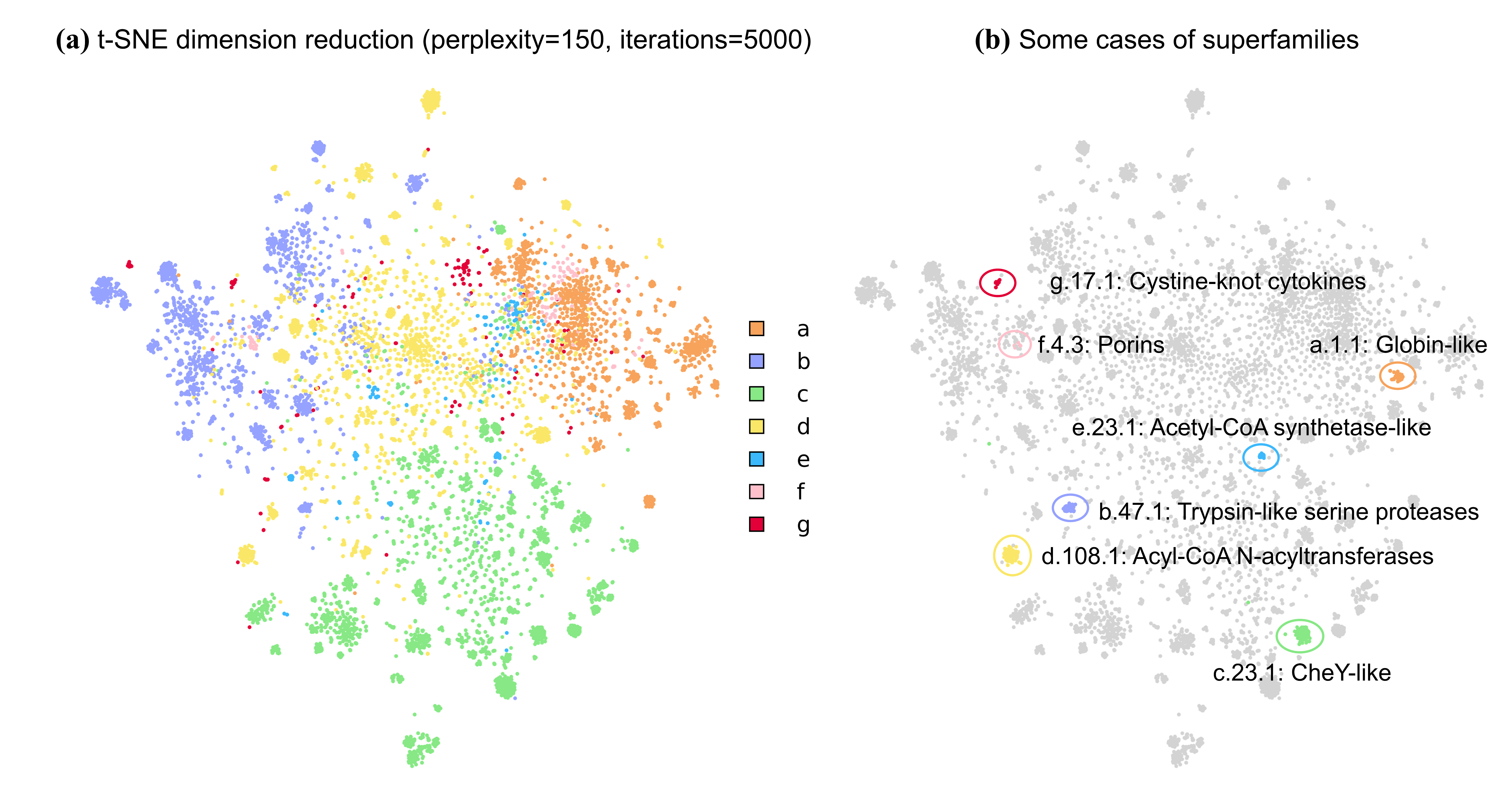}
    \caption{\textbf{Dimension reduction analysis using t-SNE.} \textbf{(a)} Visualization of the embedding learned by FoldExplorer with each color representing a class in SCOPe. \textbf{(b)} Some cases of superfamilies. The proteins of the same superfamily are clustered together.}
    \label{fig:7}
\end{figure}

Fig.~\ref{fig:7} depicts the visualization results of the embeddings learned by FoldExplorer after dimensionality reduction using t-SNE. It is evident that different structural classes are grouped into separate clusters from Fig.~\ref{fig:7}\textbf{(a)}. Moreover, we can gauge the extent of similarity between different clusters by examining their distances in the latent space. For instance, class a (all-alpha proteins) and class b (all-beta proteins) are clearly separated into two clusters, while class d (alpha and beta proteins) lies intermediary. This is because class d simultaneously possesses some characteristics of class a and also some of class b. This is in accordance with our comprehension of the protein structural space. Another example is class f (membrane and cell surface proteins and peptides), with some elements close to class a and others near class b. This is because membrane proteins encompass two major categories: alpha-helices and beta-barrels. In Fig.~\ref{fig:7}\textbf{(b)}, we selected a superfamily from each class as a representative. Proteins of the same superfamily have certain structural conservatism. It can be seen that the same superfamily is tightly clustered together, indicating that FoldExplorer has learned similarities within the superfamily and differences from other superfamilies.

All of the evidence indicates that FoldExplorer has effectively learned the spatial distribution of protein structures.

\begin{figure}[ht!]
    \centering
    \includegraphics[width=1 \columnwidth]{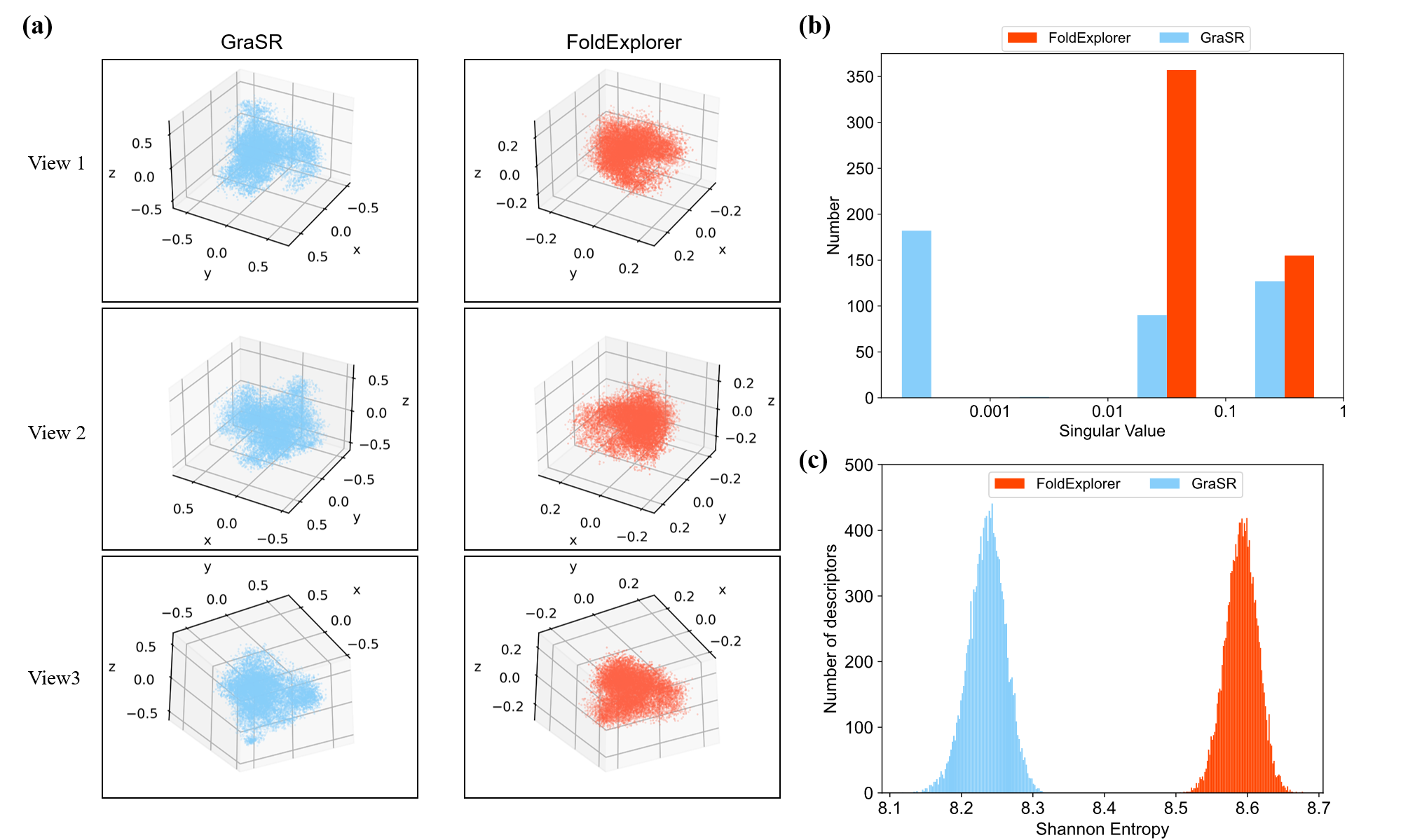}
    \caption{\textbf{Representation space compared to GraSR.} \textbf{(a)} Visualize the SCOPe descriptors learned by GraSR and FoldExplorer in 3D after dimensionality reduction using PCA. \textbf{(b)} Distribution of singular values of GraSR and FoldExplorer descriptors. \textbf{(c)} Compute and analyze the Shannon entropy distribution for each descriptor. A higher Shannon entropy indicates a greater amount of information contained.}
    \label{fig:8}
\end{figure}

\paragraph{FoldExplorer learns a more comprehensive representation of the protein structure space.}
The structural space of proteins is very complex and a high-dimensional nonlinear space. Representation-based methods aim to describe protein space through limited dimensions and use simple metrics to measure the distance between entities, which inevitably results in information loss. A effective representation method should strive to achieve minimal information loss. We compare and analyze the descriptors learned by FoldExplorer and GraSR on the SCOPe dataset,  The results are shown in Fig.~\ref{fig:8}. Fig.~\ref{fig:8}\textbf{(a)} shows the three views after PCA dimensionality reduction. Since both methods use cosine similarity as the distance metric, a space closer to a sphere is deemed more comprehensive. It is evident that the space learned by FoldExplorer is more regular and close to a sphere, while the space learned by GraSR has significant missing parts. Furthermore, we perform Singular Value Decomposition (SVD) on both representation spaces composed of descriptors and normalized the singular values (dividing all singular values by the largest one). Typically, when a singular value falls below a certain threshold (e.g., 1\% of the maximum singular value), the information in the corresponding feature is considered insignificant and can be disregarded. Fig.~\ref{fig:8}\textbf{(b)} illustrates that, despite GraSR using a 400-dimensional vector to represent the protein space, almost 200 dimensions have singular values less than 1‰ of the maximum singular value, indicating significant information loss. In contrast, FoldExplorer employs a higher 512-dimensional representation of the protein space, with singular values in all dimensions greater than 1\% of the maximum singular value, preserving ample information in each dimension. Shannon entropy, commonly used to evaluate the amount of information in a system, is also calculated for each descriptor in GraSR and FoldExplorer, and their distributions are shown in Fig.~\ref{fig:8}\textbf{(c)}. The markedly higher Shannon entropy of FoldExplorer compared to GraSR reflects that FoldExplorer has learned a more comprehensive representation of the protein structure space.

\section*{Conclusion}
As the volume of protein structure data continues to expand, the development of structure alignment and search tools has become increasingly urgent and crucial. In this study, we have introduced a novel, fast, and accurate protein structure search method FoldExplorer. FoldExplorer employs a unique approach that combines three-dimensional structural information with sequence data to encode protein structures. It leverages a multi-layer GAT network to extract information from the three-dimensional structure and simultaneously employs the Protein Large Language Model ESM to extract information from the sequence. This combination enhances the representation of protein structures. We train the encoder using a contrastive learning model MoCo to generate sequence-enhanced graph embeddings as descriptors for protein structures.

We conducted evaluations of FoldExplorer on both ranking and classification tasks, and the results indicate that FoldExplorer surpasses existing state-of-the-art methods. Furthermore, we showcased and analyzed specific instances where and why FoldExplorer outperforms Foldseek. Additionally, we visualized the protein structure latent space learned by FoldExplorer and observed a strong alignment with the original space. This suggests that FoldExplorer has, to a significant extent, learned the distribution of protein space. Furthermore, FoldExplorer is computationally efficient and achieves similar sensitivity to TM-align in one-thousandth of the computation time.

Although FoldExplorer performs quite well in protein search tasks, we aspire to further explore its capabilities in the future. Firstly, the descriptors generated by FoldExplorer can be used to assess the similarities and differences between protein structures. We can use them as pre-trained models for various downstream tasks related to protein structure, such as protein function prediction and virtual drug screening. In addition, FoldExplorer's representation of protein structure can enhance our comprehension of the protein universe, potentially leading to the discovery of new protein families and folds.

\bibliography{main}
\end{document}